# Complex networks theory for analyzing metabolic networks

ZHAO JING [1,2,4], YU HONG [2], LUO JIANHUA[1], CAO Z.W. [*2] and LI YI-XUE [* 2,3,1]

[1]Department of Biomedical Engineering, Shanghai Jiao Tong University. Shanghai 200240, China;
[2]Shanghai Center for Bioinformation and Technology, Shanghai 200235, China;
[3]Shanghai Institutes for Biological Sciences, Chinese Academy of Sciences, Shanghai 200031, China;
[4] Department of mathematics, Logistical Engineering University, Chongqing 400016, China
Correspondence should be addressed to Cao Z W and Li Yi-Xue (email:zwcao@scbit.org, yxli@sibs.ac.cn )

**Abstract**   One of the main tasks of post-genomic informatics is to systematically investigate all molecules and their interactions within a living cell so as to understand how these molecules and the interactions between them relate to the function of the organism, while networks are appropriate abstract description of all kinds of interactions. In the past few years, great achievement has been made in developing theory of complex networks for revealing the organizing principles that govern the formation and evolution of various complex biological, technological and social networks. This paper reviews the accomplishments in constructing genome-based metabolic networks and describes how the theory of complex networks is applied to analyze metabolic networks.

   The completion of the Human Genome Project started the post-genomic era. The hot topic of biological research is now shifting from the study of single genes or proteins to whole genome analyses. All kinds of "omics" technologies such as genomics, transcriptomics, proteomics and metabonomics gradually bring molecular biology into the era of system biology, and bioinformatics into that of post-genome informatics. Since genes and proteins tend to function through interacting in networks, the interaction networks must be analyzed for the purpose of studying biological functions. It is very important to analyze the biological functions in terms of the network of interacting molecules and genes, such as genetic regulatory network, signal tranduction network, protein interaction network, and metabolic network, with the aim of understanding how a biological system is organized from its individual building blocks. It represents an integration of biological knowledge from genomic information towards the understanding of the basic principles of life for biomedical applications[1].

   Metabolic process is essential for the maintenance of life. In metabolism some materials are broken down to yield energy for vital processes while other substances, necessary for life, are synthesized. The metabolic network, a characteristic complex network including all metabolites and enzyme catalyzed reactions occurring within a living cell, as well as the interactions between the reactants and enzymes, is an abstract representation of cellular metabolism. The analysis of metabolic networks can help to understand and utilize cellular metabolic process in order to promote the development of ferment technology and medicine industry. On the other hand, the topology of metabolic networks reflects the dynamics of their formation and evolution. A study of this realm may help to understand the evolutionary history of life.

---

[*]Corresponding authors:.

Actually, a lot of complex interactive nonlinear systems in the real world could be represented by networks. For a long time real world networks have been thought of as a haphazard set of nodes and connections and, thus have been described as the random networks[2]. However, in recent years, with the availability of computers and communication networks that allow us to gather and analyze data on a scale far larger than previously possible, it is found that many features of real world networks cannot be depicted by the model of random networks. The findings of small-world character by Watts and Strogatz[3] in 1998 and scale-free feature by Barabasi and Albert[4] in 1999 aroused great passion to the complex networks all over the world and marked the beginning of modern complex networks. The theory of complex networks applies methods developed in graph theory, statistics, statistical physics, and computer simulation to study the topological features, relationships between structure and function, and rules for the formation and revolution of real world networks, such as social networks (disease spread network, research paper citation network, scientist cooperation network), technology networks (the WWW network) and biological networks (food web network, metabolic network, protein interaction network). This research is already affording a new platform for studying biological networks systematically. Since Jeong et al. published their results on the topology of metabolic networks[5] in 2000, a lot of progress has been made on the structure and functionality of metabolic networks. In this review, we will present some most important methods and major results in this field.

## 1 Reconstruction of metabolic networks from genome information

Reconstruction of metabolic networks from genome data is the basis for analyzing the topology and functionality. From the annotation of genome sequence for a specific organism and the gene-enzyme relational database, all enzyme catalyzed reactions in the organism could be retrieved. The main steps are listed as follows[6]:

(i) Identifying ORFs from the genomic sequence;

(ii) predicting all the enzyme genes of this organism by sequence similarity alignment;

(iii) assigning EC numbers for the enzyme genes by querying enzyme nomenclature database such as ENZYME data bank[7] (http://expasy.hcuge.ch/);

(iv) matching EC numbers with the known reactions by querying databases concerning enzymes and reactions, such as the LIGAND/ENZYME[6] section of KEGG（ftp://ftp.genome.jp/pub/kegg/ligand/）.

This method is very useful for the reconstruction of metabolic networks, especially for new-sequenced organisms. Moreover, several metabolic reaction databases are also available including information about organism specific enzyme genes and enzymes, in which the most famous are KEGG[8,9] (http://www.genome.ad.jp/kegg; ) , EcoCyc [10,11](http://ecocyc.org) , and WIT[12]（http://wit.mcs.anl.gov/WIT/）. For example, the whole enzyme set of a specific organism, all the known reactions from all sequenced organisms and the corresponding enzymes catalyzing them could be downloaded from KEGG. Then all the reactions included in the organism could be generated.

Having determined all reactions of the organism, the metabolic network could be represented as different kinds of graph models according to the demand of individual problem. Graph is the object being studied in graph theory. A graph depicts the binary relation between the elements being studied, in which an element corresponds to a node of the graph, while the relation between two elements corresponds to a link between the two corresponding nodes. According to the different meanings of nodes, metabolic networks could be represented as 4 different kinds of graphs as follows,

1. Substrate graph: The nodes are defined as chemical compounds. If one compound can produce another through one reaction, the corresponding nodes are connected[13-18].

2. Reaction graph: The nodes correspond to reactions. There is an edge between two reactions if a compound is both a product of one reaction and a substrate of the other one[14,19].

3. Enzyme-centric graph: The nodes are defined as enzymes. Two enzymes are connected if a compound is both a product of one enzyme catalyzed reaction and a substrate of the reaction catalyzed by the other enzyme[20].

4. Substrate-enzyme bipartite graph: A bipartite graph has two sets of nodes, in which only nodes in different sets can be connected. In a substrate-enzyme bipartite graph, one set of nodes corresponds to that of compounds, while the other set of nodes corresponds to reactions. A reaction is connected with its substrates and products[5,21,22].

On the other hand, a graph can be defined in two ways, that is, undirected graph or directed graph, according to whether its edges are directed. The edge in a directed graph is called an arc. Since biochemical reactions have directions, i.e. some reactions are irreversible, metabolic networks are depicted as directed graph at most circumstances[5,13,16-19,23]. However, when studying some specific topologies of metabolic networks, the directions of reactions could be ignored and the metabolic network is thus represented as an undirected graph[14,15,21].

It is noteworthy that the current metabolites such as ATP, ADP, NADH, and $NAD^+$ are normally used as carriers for transferring electrons and certain functional groups (phosphate group, amino group, one carbon unit, methyl group, ect.) and thus take part in a lot of reactions. Including the current metabolites in the metabolic network would result in an unrealistic definition of the path length in many cases and lead to a wrong conclusion that only two steps of reactions are needed to produce pyruvate from glucose. Considering of such situation, Ma and Zeng developed a database based on the KEGG database, which excludes the connections through current metabolites and small molecules such as ATP, NADH, $H_2O$ and $NH_3$, and includes the information about the reversible reaction[16]. The metabolic network of a specific organism could be represented as a directed graph conveniently from this database.

## 2 Global structure of metabolic networks

By computational analysis of genomic data from 65 organisms in their database, Ma and Zeng discovered that the macroscopic structure of the metabolic network is organized in the form of a bow-tie, which was formerly found for the WWW network topology[17,24]. The network consists of four parts: giant strong component (GSC), substrate subset (S), product subset (P) and isolated subset (IS). The GSC is the biggest of all strongly connected components (a strongly connected component is a maximal subgraph of a directed graph such that for every pair of vertices *u, v* in the subgraph, there is a directed path from *u* to *v* and a directed path from *v* to *u*.), in which any pair of nodes have a directed path between them. *S* consists of nodes that can reach the GSC but cannot be reached from it, while P consists of nodes that are accessible from the GSC, but do not link back to it. The IS contains nodes that cannot reach, and cannot be reached from the GSC either.

Fig. 1 shows a coarse-grained graph of the metabolic network for *E.coli* drawn by us using shrunk graph approach in graph theory, in which every strongly connected component has been shrunk into a diamond node. A bigger diamond corresponds to a strong component with more nodes, while a thicker arc represents more links between the corresponding clusters. The biggest diamond in the center represents the GSC part. The bow-tie structure of the metabolic network, especially the connection topology outside the GSC part, is clearly delineated in this figure. It is noticeable that the biological information flow outside the GSC is highly branched with little cross links between the branches.

It was found that the most important cellular metabolic pathways, such as the glycolysis pathway, the TCA cycle, the amino acid synthesis pathway, and the nucleic acid synthesis pathway, are included in the GSC part. This result indicates that GSC could be the most important part in the metabolic network. It was also discovered that the reaction graph of the metabolic network shows similar bow-tie topology to that of the substrate graph [19].

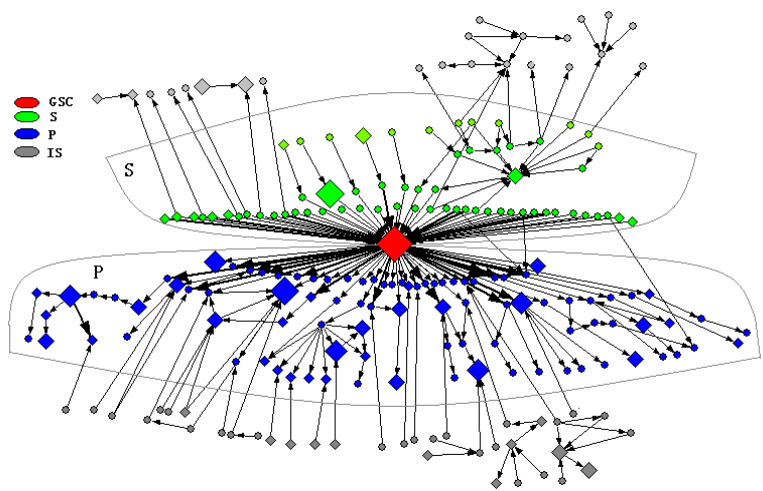

Fig. 1. Coarse-grained graph of the metabolic network for *E.coli.*

## 3  Network measurements and topological features

Each complex network presents some specific topological features that highly influence its function. The analysis and discrimination of networks therefore rely on the use of measurements capable of expressing the most relevant topological features. This section presents a survey of the most important measurements frequently used in network analysis as well as the network topologies described by them.

*3.1 Degree distribution vs scale-free networks*

For an undirected graph, the degree of a node is the number of edges connected to it. For a directed graph, the degree of a node is the sum of its out-degree and in-degree, in which the out-degree of a node is the number of links from this node to other nodes, while the in-degree is the number of links to it. The degree distribution P(k) of a network is the occurrence frequency of nodes with degree $k$, $k=1,2,\ldots$.

Since the nodes in a random graph are connected randomly, most nodes have roughly the same number of links, approximately equal to the average degree of the graph. The degree distribution of random graphs approaches Poisson distribution[2]. However, by statistical analysis of the complex networks in the real world, such as the WWW network, internet, and social network, it was found that their degree distribution follows a power law[4,24,25]. A characteristic of such network, called scaled-free network[4], is the existence of hubs, i.e. a small fraction of nodes with very high degree. This finding suggests that complex networks in the real world are not connected randomly. Fig. 2 shows the different degree distributions between random networks and scale-free networks. Jeong *et al.* studied the metabolic networks of 43 organisms in the WIT database representing all three domains of life, i.e. archae, bacterium and eukaryote. They discovered that the distribution of out-degree and in-degree of these networks approximates a power law: $p(k) \sim k^{-r}$, where the value of $r$ is about 2.2[5], indicating that metabolic networks are scale-free networks.

The degree distribution function P(k) is independent of the network's size and it therefore captures a network's intrinsic feature, which allows it to be used to classify various networks.

The origin of the power-law degree distribution was first addressed by Barabasi and Albert according to the ubiquity of scale free networks in the real world and the topological features[4]. The BA model for network evolution suggests that the scale free nature of real networks is rooted in two generic mechanisms: (1) Growth: real networks grow by the continuous addition of new nodes. (2) Preferential attachment: the likelihood connected to a node depends on the node's degree, i.e. "the rich get richer" principle. According to this model, the high-degree nodes should appear in the earlier stage of network formation. Similarly, if, early in the evolution of life, metabolic networks grew by adding new metabolites, then the most highly connected metabolites should also be the phylogenetically oldest. Glycolysis and the TCA cycle are perhaps the most ancient metabolic pathways. Wagner and Fell listed the thirteen hubs in the substrate graph of the metabolic network for *E. coli*. This list includes many intermediates in glycolysis and the TCA cycle, such as acetyl CoA, pyruvate, 3-phosphoglycerate, succinate, and 2-oxoglutarate, along with the amino acids thought to be used earliest, such as glutamine, glutamate, aspartate, and serine[14]. This result indicates that purely topological analysis of metabolic networks could reveal the evolution history and design principle of the networks to some extent.

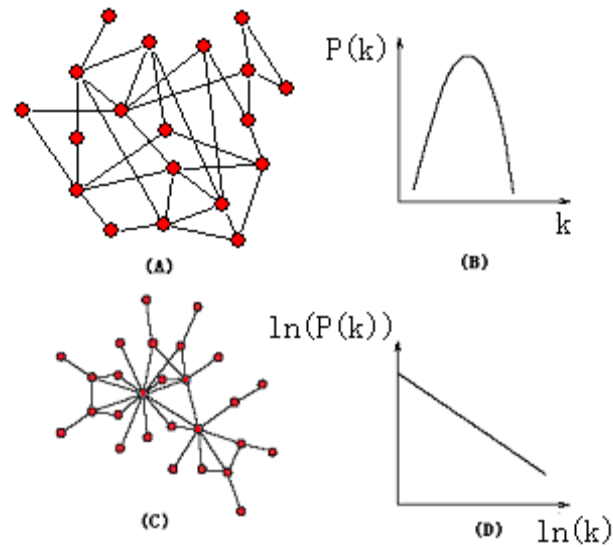

Fig. 2. Degree distribution of random networks and scale free networks.
(a) A random network;   (b) Degree distribution of the random network.
(c) A scale free network;   (d) Log-log plot of the degree distribution for the scale free network.

*3.2 Clustering coefficient vs Hierarchical modular networks*

In an undirected graph, the clustering coefficient of node *v* measures the extent of its neighbours to form a clique. The mathematical definition is as follows:

$$CC(v) = \frac{2|N(v)|}{d(v)(d(v)-1)},$$

where $|N(v)|$ denotes the number of links between neighbours of node *v*, $d(v)$ is the degree of node *v*. The value of *CC (v)* is between 0 and 1. According to the definition, *CC(v)* equals 1 if any pair of neighbours of *v* have direct links, while *CC(v)* is 0 if no direct links exist between neighbours of *v*.

Function *C(k)* is defined as the average clustering coefficient of all nodes with *k* links. Ravasz *et al.* calculated function *C(k)* for the metabolic networks of 43 organisms and found that the distribution of *C(k)* obeys a power law: $C(k) \sim k^{-1}$, which is the same as that of the hierarchical modular networks constructed by them. Similar to degree distribution, the distribution of *C(k)* is independent of the network size and thus could be used to distinguish different networks. Therefore, metabolic networks are believed to be organized as many small compacted modules being connected in a hierarchical manner to form larger modules, which in turn are integrated into even larger modules. Ravasz et al. proposed an algorithm to decompose a metabolic network into small modules and found that most topological modules contain reactions mostly from one major pathway, indicating that the modules are functional modules with relatively independent biological functions[15,26]. From topological view, this result proves that metabolic networks present modular organizations, like other biological systems[27,28].

*3.3 Mean path length vs small-world networks*

The distance of two nodes in a network can be measured by the length of the shortest path, which is one of the paths connecting these nodes with minimum edges or arcs. The diameter of a network is the length of the longest short path in the network, while the mean path length is the average length of the shortest paths between any pair of nodes in the network. These two parameters can measure the average communication speed between the nodes of the network.

Another important feature of complex networks is "small-world" property, i.e. all nodes can be reached from the others through a small number of nodes, being analogous to the small-world phenomenon popularly known as six degrees of separation[3]. Jeong *et al.* calculated the mean path length of the metabolic networks for 43 organisms and found that almost all the networks have a mean path length of 3.2, suggesting that metabolic networks are small-world networks. This result means that most of the metabolites can be converted to each other in about 3 steps, which is not accordant with the real situation of cellular biochemical reactions. The reason is that the networks reconstructed by Jeong *et al.* included the current metabolites such as ATP and NADH, thus resulting in an unrealistic definition of the path length. Ma and Zeng reconstructed the metabolic networks of 80 organisms by excluding the current metabolites and small molecules and then found that the average values of the mean path length for eukaryote, bacteria and archaea are 9.57, 8.50 and 7.22, respectively[16]. Though these values are much bigger than that obtained by Jeong *et al.*, they are still at the logarithmic scale of the network size, also suggesting the small-world feature of metabolic networks. The small-world nature of metabolic networks may allow a metabolism to react rapidly to perturbations in enzyme or metabolite concentrations; thus the cell may react quickly to changes of the surroundings[14].

## 4 Decomposition of metabolic networks

Some pathway analysis methods such as elementary flux modes (EFMs) and extreme pathways (EPs)[13,29,30] have been shown to be useful for functional analysis of metabolic networks. However, these methods often meet the problem of combinatorial explosion when applied to genome-scale metabolic networks. Therefore, it is necessary to break the metabolic network up into relatively functional independent sub-networks before performing functional analysis using these methods.

Usually, the decomposition of metabolic networks is performed on an intuitive basis. It is a common practice for biologists to discuss metabolism in terms of distinct biochemical pathways, such as sugar metabolism, lipid metabolism, and amino acid metabolism. Schilling and Palsson proposed to subdivide the network into relatively isolated clusters of reactions, according to intuitive biological criteria[31]. On the other hand, considering that the topology of network is a reflection of its function and the metabolic network has been revealed to be organized in a hierarchically modular manner, more algorithms purely based on the network topology have been proposed. These algorithms split the metabolic network into individual topological modules, in which network nodes are joined together tightly, between which there are only looser connections. This kind of decomposition can also help to understand the organization principle of complex biological system.

Based on the scale-free feature of metabolic networks, Schuster *et al.* proposed a decomposition algorithm by removing the high-degree nodes (hubs) whose degree is bigger than a threshold. Using this method, they broke the metabolic network of a parasitic bacterium mycoplasma pneumoniae up into 19 relatively isolated biological modules, such as arginine degradation, the tetrahydrofolate system, and nucleotide metabolism[32].

Guimera and Amaral addressed network decomposition as an optimization problem and tried to find the best decomposition. Using simulated annealing method, they identified modules by maximizing the network's modularity parameter so that there are as many as within-module links and as few as possible between-module links. The modules obtain by this algorithm are biologically functional[18].

The two algorithms described above belong to nonhierarchical clustering method, which only focuses on decomposing the network but does not consider the hierarchical connection between modules. Another kind of decomposition uses hierarchical clustering method, which can show the hierarchical modularity of the network by a rooted tree. This kind of method starts with a properly defined similarity index or dissimilarity index, which signifies the extent to which two nodes would like in the same cluster, and then attempts to divide the network nodes into clusters based on this index using agglomerative or divisive method. Agglomerative method starts off with each node being its own cluster. At each step, it combines the two most similar clusters to form a new larger cluster until all nodes have been combined into one cluster. Divisive method begins with one cluster including all the nodes, and attempts to find the splitting point at which two clusters are as dissimilar as possible. The process continues until all clusters have size one. The main discrimination of different hierarchical clustering methods is that they use their own similarity index or dissimilarity index. Several hierarchical clustering methods proposed to decompose the metabolic network are surveyed as follows.

Ravasz *et al.* first reduced the metabolic network by the biochemical and topological reduction process and treated the network as an undirected graph by ignoring the reaction directions. Then they decomposed the condensed network by agglomerative method[15]. A similar index between node *i* and *j* is defined as the topological overlap as follows,

$$O_T(i,j) = \frac{J_n(i,j)}{\min(k_i, k_j)},$$

where $J_n(i,j)$ denotes the number of nodes to which both *i* and *j* are linked ( plus 1 if there is a direct link between *i* and *j* ); $k_i$, $k_j$ is the degree of *i* and *j*, respectively.

Ma and Zeng proposed an algorithm based on the bow-tie structure of metabolic networks. They first decompose the GSC part of the network and then expand the partition to the whole network using a majority rule[19]. The dissimilarity index between node *i* and *j* of GSC part is defined as the smaller one of the two path lengths between this pair of nodes,

$$dissimilarity(i,j) = \min(d(i,j), d(j,i)),$$

where $d(i, j)$ is the number of arcs in the shortest directed path from *i* to *j* (in general, $d(i,j) \neq d(j,i)$).

Holme *et al.* denoted the metabolic network by a substrate-enzyme bipartite graph and defined the dissimilarity index as the betweenness centrality of the reaction node *r*,

$$C_B(r) = \frac{1}{k_{in}(r)} \sum_{s \neq t} \frac{\sigma_r(s,t)}{\sigma(s,t)},$$

where $\sigma_r(s,t)$ is the number of shortest paths between *s* and *t* that passes through *r*, $\sigma(s,t)$ is the total number of shortest paths between *s* and *t*, $k_{in}(r)$ is the in-degree of node *r*. Divisive method was used to decompose the network into modules by successively removing reaction nodes of high betweenness centrality[33].

## 5  Robustness of metabolic networks

Robustness, a property that allows a system to maintain its functions under certain external and internal perturbations, is a ubiquitously observed nature of biological systems. For example, microbes can still grow under the circumstance of knock-out mutation, and sometimes even keep the same growth rate as their wild types. Studying the interplay between the topology and robustness of metabolic networks may help to understand the functional organization principle of cells and could have important implication for disease studies[34,35] and drug target identifications[36,37].

The robustness of the metabolic network with respect to specific enzymes can be qualitatively estimated by the preservation or decay of the network after removal of the nodes or edges corresponding to these enzymes. The following topological features of metabolic networks may ensure the robustness of metabolism.

1. Modularity: Modularity contributes to the robustness of networks by decreasing the cross talk between different functional modules, confining perturbations and damages to separable parts and preventing deleterious effects from spreading to the whole system[38,39].

2. Bow-tie structure: The GSC part in the bow-tie structure of the metabolic network forms a robust conserved core because it is the most tightly connected part of the network and there are multiple routes between any pair of nodes within the GSC. Moreover, the GSC includes the most important pathways such as glycolysis pathway and the TCA cycle. On the other hand, the S and P part are connected with the GSC by the branched multiple-input and multiple-output way. Such connecting pattern provides an advantage in generating coordinated response to various stimuli and increases the robustness of the whole system[38].

3. Scale-free topology: The key feature of scale-free networks is the high degree of error tolerance; that is, the ability of their nodes to communicate is unaffected by the failure of some randomly chosen nodes[40]. Thus the scale-free nature of metabolic networks indicates its high resistivity to random perturbations and could explain why some enzyme dysfunction at the metabolic level is without substantial phenotypic effect[5]. On the other hand, scale-free networks are extremely vulnerable to attacks, i.e. the removal of a few hub nodes which play a crucial role in maintaining a network's connectivity will destroy the whole network[40]. It is interesting that Mahadevan and Palsson showed that low-degree nodes are almost as likely to be critical to the overall network functions as high-degree nodes[41] recently, by calculating the number of lethal reactions among all the reactions connected to every metabolite in the substrate graph of metabolic networks. Samal et al farther found that almost all essential reactions are tagged by low-degree metabolites[42]. These findings are different from the proposed error tolerance and attack vulnerability properties of metabolic networks[5,40]. The possible reason is that the hub nodes being deleted in the previous study[5,40] correspond to metabolites. However, a metabolite in metabolic networks cannot be deleted by genetic techniques, but the enzymes that catalyze reactions can[41]. This indicates that the interplay between the structure and function of real networks is so complex that it is difficult to draw conclusion from only one aspect of topology.

Palumbo *et al.* studied the relation of enzyme mutations in Saccharomyces cerevisiae metabolic network and their lethality. They found that the deletion of the edges corresponding to essential enzymes would break the whole network into several separated sub-networks, or interrupt the full connectivity in a strongly connected component[43]. Lemke *et al.* related the extent of the topological damage generated in the *E.coli* metabolic network by deleting an enzyme to the essentiality of the enzyme determined by experiments. The damage measurement $d$ of an enzyme to the network is defined as the number of metabolites whose production is limited by the absence of the enzyme. They found that the $d$ value of most enzymes is smaller than 5, while only 9% of enzymes have $d$ value no less than 5, more than half of which are essential enzymes[22]. This group also used their method to determine the most important enzymes of the metabolic network of mycoplasmas and discovered that the genomes of six mycoplasmas shared predicted important enzymes[44]. All of these methods could have practical application in the target identification for gene engineering or for drugs.

## 6 Conclusions and prospects

In the past few years, significant advances have been achieved in the research of metabolic networks and they help to understand cellular metabolism and functions. The topological features of metabolic networks, such as scale-free, small-world and hierarchical modularity, are also shared by genetic regulatory networks, protein interaction networks, and other biological networks[45,46]. These common characteristics help us to comprehend biological systems from a higher abstract view and to reveal new biological features.

According to the hierarchical modular feature of cellular networks, the cellular organization system could be described as a life's complexity pyramid[47]: The bottom shows the various molecular components of the cell—genes, RNAs, proteins, and metabolites; these elementary building blocks organize themselves into small recurrent patterns, i.e. pathways in metabolism and motifs in genetic regulatory networks (level 2); in turn, motifs and pathways are then integrated to form functional modules responsible for discrete cellular functions (level 3); these modules are nested in a hierarchical way and define as the cell's large-scale functional organization—the top of the pyramid. From the bottom to the top shows transition from the particular to the universal. Although the individual components are unique to a given organism, the topologic properties of cellular networks share similarities. From the top to the bottom there is a transfer from universality to organism specificity. Although more detailed studies are required in order to understand the topological relations among pathways, motifs, modules and the whole network, this viewpoint offers a rough roadmap for understanding complex biological systems. It is one of the most important tasks in the post-genomic era to get a detailed comprehension of the interplay between structure and function of biological systems and to identify the specific structural and functional modules with relation to diseases and cancers.

Although several advances have been achieved in the topology of metabolic networks, little has been known about its formation principle, that is, the role of evolution and selection in shaping cellular networks. For example, most complex networks are scale-free networks, in which a few hubs are known to play a key role in keeping the network together. The connection pattern of hubs has important effect on network behavior. Several measurements suggest that in cellular and technological networks the hubs tend to be connected with less connected nodes (disassortative mixing), while in social networks the hubs have the tendency to be linked together[48] (assortative mixing). We should further study the evolutionary mechanism that shapes cellular networks into the opposite linkage pattern with that of social networks and the impact of such linkage pattern on the function of networks.

The study of cellular networks is still in its infancy[46]. Further progress is expected about the topology, function and evolutionary mechanism of cellular networks. On the other hand, since information about metabolic kinetic parameters is rather incomplete, most studies have so far focused on the static features of networks. It is expected that more advances in the dynamic aspects of various cellular networks could be achieved along with the progress of genomics, proteomic and metabonomics, and the accumulation of cellular kinetic information. These studies will certainly be helpful for the modeling and simulation of complex biological systems and will promote important biomedical applications.

**Acknowledgements**   This work was supported by the National Science and Technology Key Programs of China（Grant No: 2004BA711A21）, the State Key Program of Basic Research of China（Grant No: 2004CB720103）, the National Natural Science Foundation of China(Grant No: 30500107) and the Key Program of Basic Research of Shanghai(Grant No: 04QMX1450，04DZ19850，04DZ14005).